\documentclass[11pt,pre,epsfig,address,onecolumn]{revtex4}%
\usepackage[english]{babel}
\usepackage[latin1]{inputenc}
\usepackage[dvips]{graphicx}
\usepackage{amsmath}
\usepackage{amssymb}
\usepackage{epsfig}
\usepackage{array}
\usepackage{amsfonts}
\usepackage{graphicx}
\usepackage{color}
\usepackage{subfigure}%
\newcommand{\be}{\begin{equation}}
\newcommand{\ee}{\end{equation}}
\newcommand{\bel}[1]{\begin{equation}\label{#1}}
\newcommand{\bea}{\begin{eqnarray}}
\newcommand{\eea}{\end{eqnarray}}
\newcommand{\balign}{\begin{align}}
\newcommand{\ealign}{\end{align}}
\newcommand{\ba}{\begin{array}}
\newcommand{\ea}{\end{array}}
\newcommand{\bfig}{\begin{figure}}
\newcommand{\efig}{\end{figure}}

\begin{document}
\title{Boundary-driven phase transitions in open two-species driven systems with an
umbilic point}
\author{Vladislav Popkov$^{1,2}$ }
\affiliation{$^{1}$ Max Planck Institute for Complex Systems, N\"othnitzer Straße 38, 01187
Dresden, Germany}
\affiliation{$^{2}$ Dipartimento di Fisica e Astronomia, Universit\`a di Firenze, via G.
Sansone 1, 50019 Sesto Fiorentino, Italy}
\date{\today}

\begin{abstract}
Different phases in open driven systems are governed by either shocks or
rarefaction waves. A presence of an isolated umbilic point in bidirectional
systems of interacting particles stabilizes an unusual large scale excitation,
an umbilic shock (U-shock). We show that in open systems the U-shock governs a
large portion of phase space, and drives a new discontinuous transition
between the two rarefaction-controlled phases. This is in contrast with
strictly hyperbolic case where such a transition is always continuous. Also,
we describe another robust phase which takes place of the phase governed by
the U-shock, if the umbilic point is not isolated.

\end{abstract}
\maketitle




\section{Introduction}

\label{sec::Introduction}

Many intrinsically nonequilibrium phenomena can be observed already in
simplest systems of driven diffusing particles \cite{Ligg99,Schu01}, which are
paradigmatic models of systems far from equilibrium and find a wide range of
applications in biological, social and physical contexts
\cite{Muka00,Evan05,Scha10}. Driving forces due to bulk fields or boundary
gradients lead to steady state currents that invalidate the condition of
detailed balance and give rise to remarkable features which have no
equilibrium counterparts, such as boundary driven phase transitions,
spontaneous symmetry breaking and hysteresis in one spatial dimension. Models
with two or more conserved species of particles exhibit particularly rich
behaviour \cite{Schu03}.

The evolution of driven systems on large spatio-temporal scales is governed by
two fundamental types of excitations: shocks, which carry discontinuities, and
rarefaction waves, which are continuous self-similar solutions of the
hydrodynamic limit equations \cite{Kipn99}. Various properties of the
fundamental excitations like stability, speed and morphology are determined by
a scalar or vector function which relates steady macroscopic currents to
average particle densities, the so-called current density relation. The
topology of the current-density function such as the number of extrema and
saddle points determines qualitative features of the large scale dynamics and
in particular the number and character of the different stationary phases and
phase transitions that one can observe in the underlying microscopic model
\cite{Popk99,PopkovCambridge,PopkovHierarchy11}. In this way microscopic
details of local particle interactions are largely irrelevant as long as they
produce a certain type of a current density relation.

It was noted \cite{UshockPRE2012} that bidirectional particle systems, in
which bulk hopping rates of oppositely moving interacting species possess the
left-right symmetry, have a special property: their current-density function
has an umbilic point. A generic umbilic point is a point on a current-density
surface where the two characteristic velocities coincide, which breaks the
usual assumption of strict hyperbolicity \cite{ChenKan95}. For bidirectional
systems with left-right symmetry, both characteristic velocities vanish at the
umbilic point which can be isolated or not, depending on a strength of an
interaction between the species. An isolated umbilic point, in an open system
under maximal feeding regime (a regime where particles enter and exit the
system freely) stabilize a large scale excitation reminiscent of a shock wave,
but which should be unstable according to usual shock stability criteria
\cite{Lax73},\cite{Lax2006}. \ The new excitation was called an umbilic shock,
or a U-shock, and studied on microscopic scale \cite{UshockPRE2012}.

In present article we determine a domain on the phase diagram which is
controlled by a U-shock, and identify a boundary driven phase transition that
it governs. Boundary-driven phase transitions in driven systems, which are
caused by adiabatical changes of boundary conditions have no equilibrium
counterparts \cite{Krug91},\cite{Popk99},\cite{InfiniteReflections}. They may
be continuous or discontinuous, depending on whether the order parameter
changes across the transition in a continuous or discontinuous way. We show
that a U-shock governs a discontinuous phase transition from one
rarefaction-wave controlled state to another. Such a transition in a usually
considered strictly hyperbolic systems (without an umbilic point) is always
continuous \cite{PopkovHierarchy11}. If the umbilic point is not isolated, the
U-shock is no longer stable, but on its place we find another robust phase, a
homogeneous bulk state with densities matching the umbilic point. This state,
which we call the umbilic state, has the same stability domain as the U-shock
phase. A boundary driven phase transition leading to the umbilic state occurs
via a continuous transition. We expect our resuts to be generally valid for
any system with an umbilic point; however for definiteness we consider a model
for which the current-density relation is known exactly.

The plan of the paper is the following: In Sec.
\ref{sec::The bidirectional model} we introduce our model. In Sec.
\ref{sec::Splitting} we discuss splitting of the physical region according to
signs of characteristic velocities, and review the U-shock and the umbilic
phase. In Sec.\ref{sec::Boundary-driven} we describe phase transitions from
and to the phases controlled by umbilic point, along a trajectory, where
boundary rates are changed adiabatically. There, we also describe the domain
of stability of the umbilic point- controlled phases. We finish with
conclusions and perspectives. Appendices contain necessary technical details.

\section{The bidirectional model with boundary reservoirs}

\label{sec::The bidirectional model}

\begin{figure}[ptb]
\centerline{\scalebox{0.6}{\includegraphics{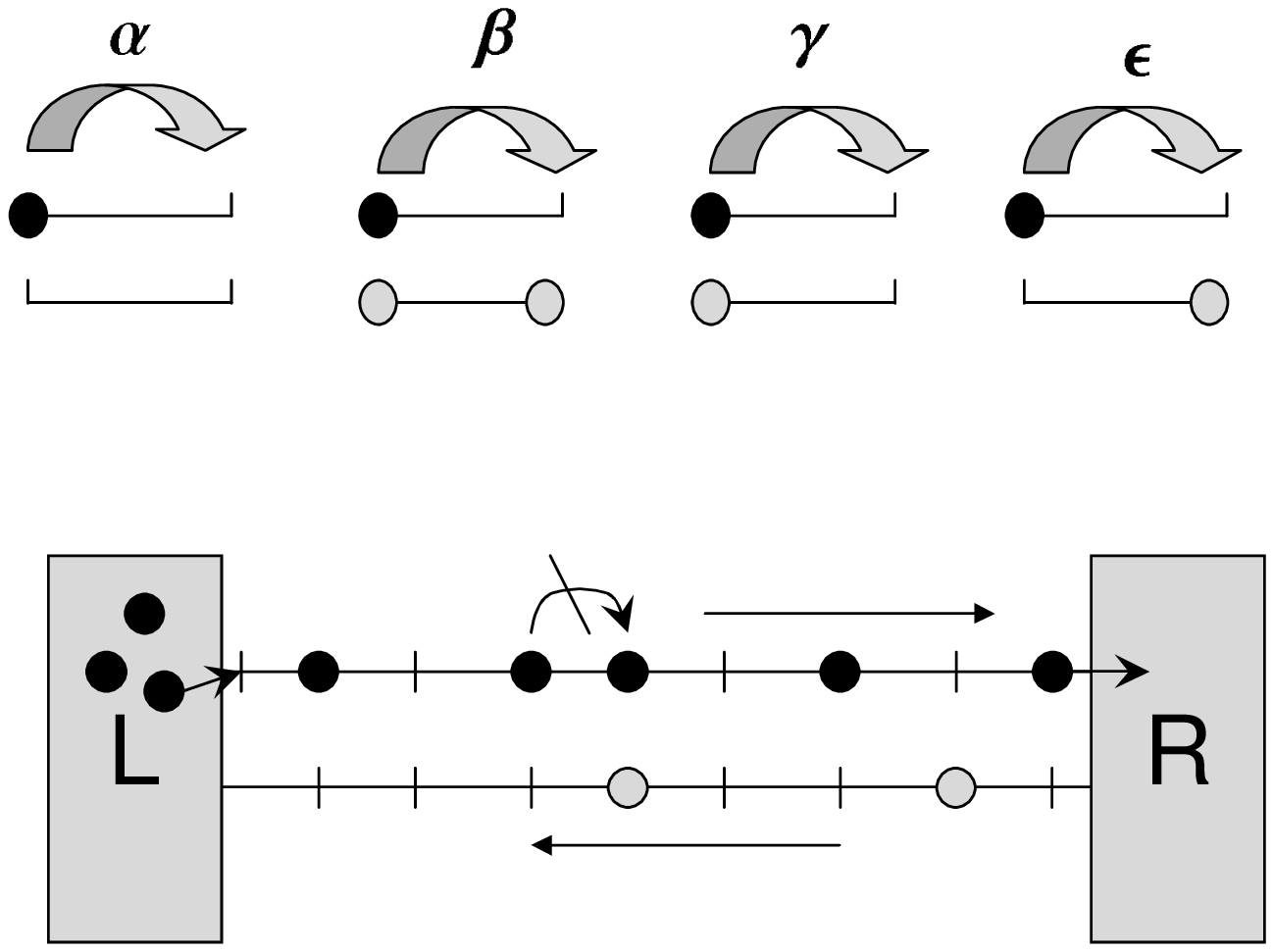}}
}\caption{Bidirectional two-chain model. For solvability, the rates must
satisfy $\alpha=\beta=\varepsilon=1$, $\gamma= \mathrm{e}^{\nu}$ where $\nu$
is the interchain interaction constant \cite{GunterSlava_StatPhys03}. Coupling
to boundary reservoirs is indicated by boxes marked L (the left reservoir) and
R (the right reservoir).}%
\label{Fig_BidirectionalModel}%
\end{figure}

Our model describes particles with repulsive hard-core interaction which hop
unidirectionally along two chains of $N$ sites: One chain for right-hopping
particles and another chain for left-hopping particles. At each instant of
time the system is fully described by occupation numbers $n_{k}\in\{0,1\}$
(for the right movers) and $m_{k}\in\{0,1\}$ (for the left-movers). A
right-moving particle at site $k$ can hop to its neighbouring site $k+1$
provided it is empty, with a rate that depends on the occupancies at sites
$k,k+1$ on the adjacent chain, see Fig.\ref{Fig_BidirectionalModel}. E.g. a
particle hops with rate $\beta$ if the adjacent sites are both occupied, etc.
For clarity of presentation and analytic simplification we shall keep only one
rate $\gamma=\mathrm{e}^{\nu}$ different from others, setting all remaining
rates to $1$,%
\begin{equation}
\alpha=\beta=\varepsilon=1,\quad\gamma=\mathrm{e}^{\nu} \label{rates}%
\end{equation}
Then the parameter
\begin{equation}
Q=\gamma-1 \label{InteractionQ}%
\end{equation}
which ranges from $\ -1$ to $\infty$, measures the interaction strength
between the left- and right-moving species. For $Q=0$ the model reduces to two
independently running totally asymmetric exclusion processes \cite{Schu93}%
,\cite{Derr93}.

The bulk dynamics of particles (see Fig.\ref{Fig_BidirectionalModel}) is
complemented with boundary conditions: we consider open boundaries where at
the left end of the chain a right mover can enter the chain and it can leave
it at the right end. Left movers are hopping to the left with the same dynamic
rules. The boundary hopping rates are chosen so as to correspond to particle
reservoirs with effective densities of right- and left movers $u_{L} $,$v_{L}$
at the left boundary, and $u_{R}$,$v_{R}$ at the right boundary, see Appendix
A. Note that generically $u_{L}\neq v_{L},u_{R}\neq v_{R}$ , so that while the
bulk dynamics is left-right symmetric, the entrance and exit rates for
different species are not. After a certain transition period, the system will
approach a stationary state, characteristics of which (the average flux, the
density profile, the correlations) do not depend on time.

For our dynamical Monte-Carlo simulation we choose the following random
sequential update procedure. For a chain of length $N$, i.e. a system of $2N$
sites (numbered $i=1,2,...N$ for right movers and $i=N+1,N+2,...2N$ for left
movers) one Monte-Carlo step consists of $2N+2$ uniform drawings of an integer
random number $w$ in the range $0\leq w\leq2N+1$. If $0\leq w\leq N$, the
configuration of right movers is updated. If $w=0$, and the left boundary site
$i=1$ is empty, we fill it with a particle with a rate parametrized by
boundary reservoir densities $u_{L},v_{L}$, see Appendix
\ref{Appendix::Boundary rates}. If $w=N$ and the respective site contains a
particle, we remove it with a rate parametrized by boundary reservoir
densities $u_{R},v_{R}$, see Appendix \ref{Appendix::Boundary rates}. For
intermediate $0<w<N$, if site $w$ contains a particle, a hopping is performed
on the right neighbouring site with given rates (\ref{rates}), provided it was
empty. The update of the left movers is done analogously. We start from an
empty lattice and after a transient time we measure site occupancies
$n_{k},m_{k}$, and take averages over many Monte Carlo steps and many
histories. We choose a system size up to $L=500$ sites in each chain. The
transient time for $L=500$ is $10^{5}$ Monte Carlo steps, and averaging over
up to $10$ histories is done.

In contrast to our study \cite{UshockPRE2012} which was focused on microscopic
features, here we focus on large-scale hydrodynamic behaviour of an open
system with an umbilic point. To this end, we also use an improved meanfield
approach, described in Appendix \ref{Appendix::Meanfield equations}. Results
obtained by the stochastic approach and the meanfield approach agree well,
both for system dynamics and for steady state global averages, due to
product-measure steady state property on a ring (\ref{ProductMeasure}).

\section{Splitting of physical regions according to signs of characteristic
velocities}

\label{sec::Splitting}

Characteristic velocities $c_{1}(u,v)$ and$\ c_{2}(u,v)$ are velocities with
which infinitesimal perturbations are propagating, on top of a stationary
homogeneous background with average densities of right- and left-moving
particles $u$ and $v$. As such, they play a fundamental role in the stability
of large scale excitations \cite{GunterSlava_StatPhys03}.

The characteristic velocities can be obtained by solving an eigenvalue problem
for a flux Jacobian $(Dj)\Psi_{k}=c_{k}\Psi_{k}$ where%
\begin{equation}
(Dj)=%
\begin{pmatrix}
\frac{\partial j_{1}}{\partial u} & \frac{\partial j_{1}}{\partial v}\\
\frac{\partial j_{2}}{\partial u} & \frac{\partial j_{2}}{\partial v}%
\end{pmatrix}
.\label{Jacobian}%
\end{equation}

For our model (\ref{rates}) the particle currents $j_{1}$ and $j_{2}$ can be
obtained analytically, see \cite{UshockPRE2012} for details, and are given by
\begin{align}
j_{1}(u,v) &  =u(1-u)+Q\Omega_{11}(u,v)\Omega_{00}(u,v)\label{Ju}\\
j_{2}(u,v) &  =-j_{1}(v,u)=-v(1-v)-Q\Omega_{11}(v,u)\Omega_{00}%
(v,u),\label{Jv}%
\end{align}
where $\Omega_{11}$ and $\Omega_{00}$ are are stationary probabilities to have
two adjacent particles and two adjacent holes,
\begin{align}
\Omega_{11} &  =\frac{(u+v-1)Q-1+\sqrt{\left(  (u+v-1)Q-1\right)  ^{2}+4Quv}%
}{2Q}\label{Omega11}\\
\Omega_{00} &  =1-u-v+\Omega_{11}.\nonumber
\end{align}

\begin{figure}[ptbh]
\begin{center}
\subfigure[\label{Fig_Split_NonCrit}] {\includegraphics[width=0.45\textwidth]{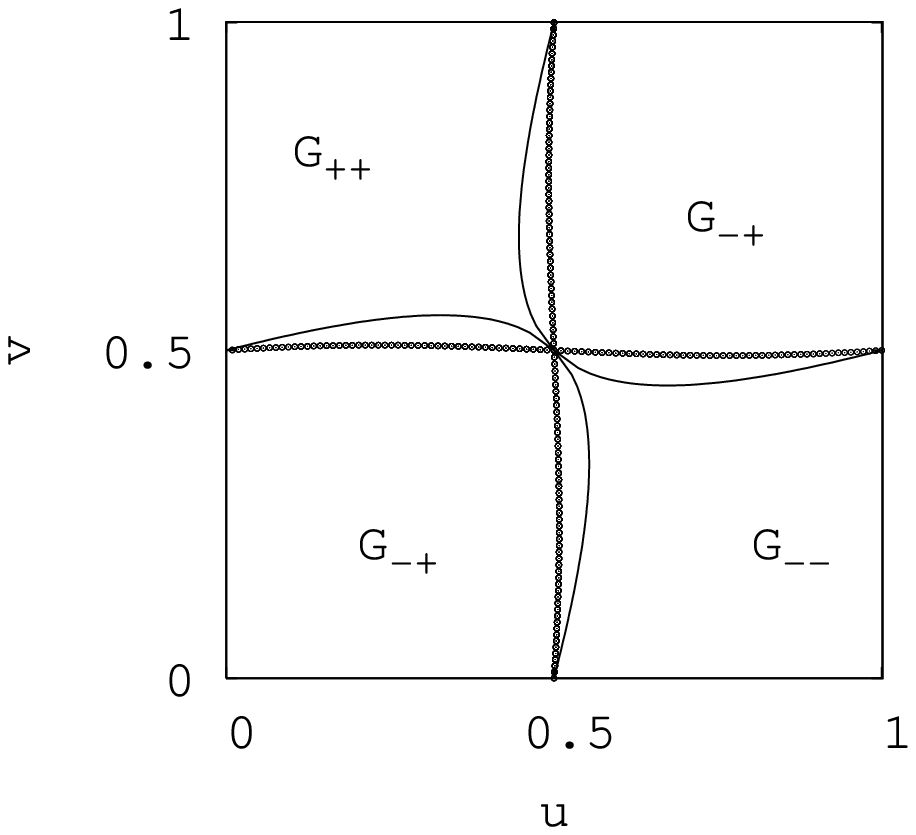}}
\qquad
\subfigure[\label{Fig_Qcrit}] {\includegraphics[width=0.45\textwidth]{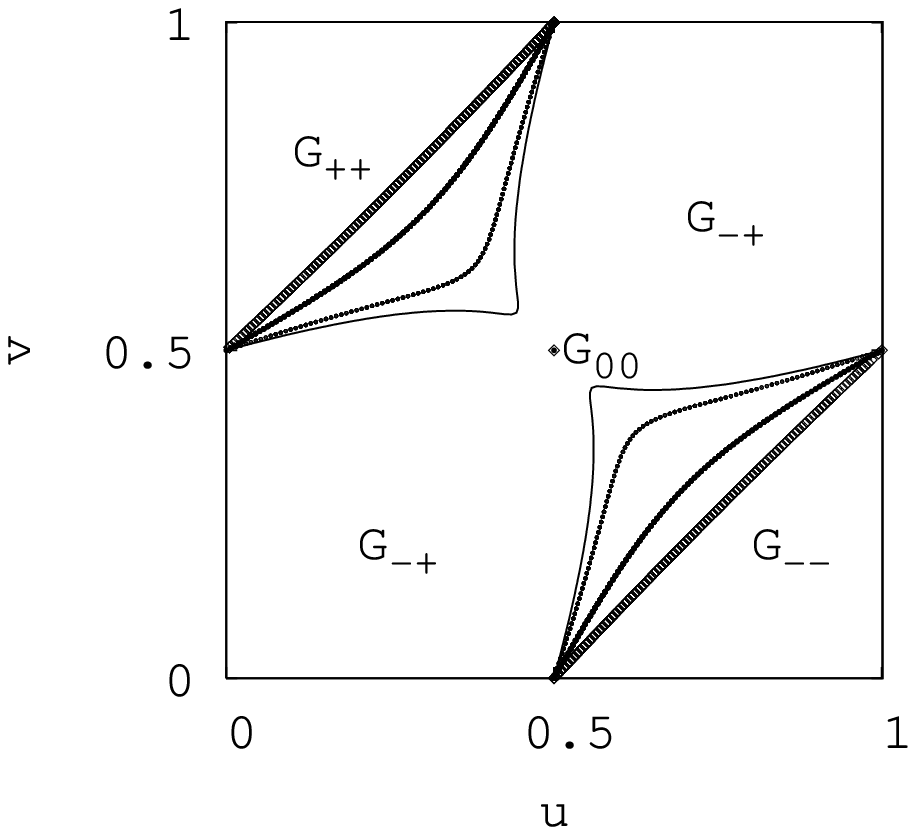}}
\end{center}
\caption{Splitting of the physical region into domains $G_{++},G_{--},G_{-+}$
according to signs of characteristic velocities, for different $Q$. Boundaries
between the domains, on which one characteristic velocity vanishes
$c_{i}(u,v)=0$, are marked by lines. Point $u=v=1/2$ is an umbilic point where
$c_{1}=c_{2}=0$ for any value of $Q$. ( \textbf{Panel (a)}: $Q \geq Q_{crit}$.
Thick and thin lines stand respectively for $Q=-0.5,-0.75$. Umbilic pont is
situated at the crossing of two curves. \textbf{Panel (b)}: $Q \leq Q_{crit}$.
Lines, in order of increasing thickness, correspond to $Q=-0.76,
-0.8,-0.9,-0.99$. Umbilic point, marked by $G_{00}$, is an isolated point. }%
\label{FigSplitting}%
\end{figure}

Due to the hardcore exclusion, the average densities of the right and
left-moving particles may only take values between $0$ and $1$. The whole
physical region of $0\leq u,v\leq1$ is then splitted into regions
$G_{\sigma\tau}$with different signs of characteristic velocities, which is
illustrated in Fig.\ref{FigSplitting}. Subscripts $\sigma,\tau$ denote signs
the characteristic speeds $c_{1}$ and $c_{2}$, i.e. $\sigma=0,+,-$ correspond
to $c_{1}=0,c_{1}>0,c_{1}<0$, and similarly for $\tau$. E.g. we name by
$G_{-+}$ a region on $u-v$ plane where $c_{1}(u,v)<0$ and $c_{2}(u,v)>0$. Note
that the characteristic speeds are numerated in increasing order, $c_{1}%
<c_{2}$.

As we can see in Fig.\ref{FigSplitting}, the splitting contains a special
point, an umbilic point, $u^{\ast}=v^{\ast}=\frac{1}{2}$, where characteristic
velocities both vanish, $c_{1}(u^{\ast},v^{\ast})=c_{2}(u^{\ast},v^{\ast})=0$,
for any value of $Q$, as can be straighforwardly verified from (\ref{Ju}%
),(\ref{Jv}). For $Q>Q_{crit}=-\frac{3}{4}$, the umbilic point is a crossing
point of the curves $c_{1}(u,v)=0$ and $c_{2}(u,v)=0$, see
Fig.\ref{FigSplitting}(a). The respective current-density surfaces
$j_{k}(u,v,Q)$ have a regular convex topology. For $Q<Q_{crit}$,
current-density surfaces $j_{k}(u,v,Q)$ develop a saddle point, and the
umbilic point becomes an isolated point, see Fig.\ref{FigSplitting}(b).

Characteristic speeds determine stability of large scale excitations in our
system, described on the macroscopic scale by a system of conservation laws
for coarse- grained densities $u(x,t),v(x,t)$
\begin{align}
\partial_{t}u+\partial_{x}j_{1}(u,v)  &  =0\label{PDE}\\
\partial_{t}v+\partial_{x}j_{2}(u,v)  &  =0,\nonumber
\end{align}
$j_{1}$ and $j_{2}$ being steady particle currents \cite{DerivationHydro},
completemented with boundary conditions
\begin{align*}
u(0,t)  &  =u_{L};\ u(1,t)=u_{R},\\
v(0,t)  &  =v_{L};\text{ \ }v(1,t)=v_{R}.
\end{align*}

A commonly made assumption about the flux functions $j_{1},j_{2}$, called
strict hyperbolicity, reads: the characteristic speeds are different
$c_{1}(u,v)$ $\neq c_{2}(u,v)$ for all $u,v$. Strictly hyperbolic systems have
only two types of fundamental solutions: shocks and rarefaction waves
\cite{Lax2006}. Presence of an umbilic point ruins strict hyperbolicity and
results in appearance of novel excitations listed below.

\begin{figure}[ptbh]
\begin{center}
\subfigure[\label{Fig_UprofileQm9}] {\includegraphics[width=0.45\textwidth]{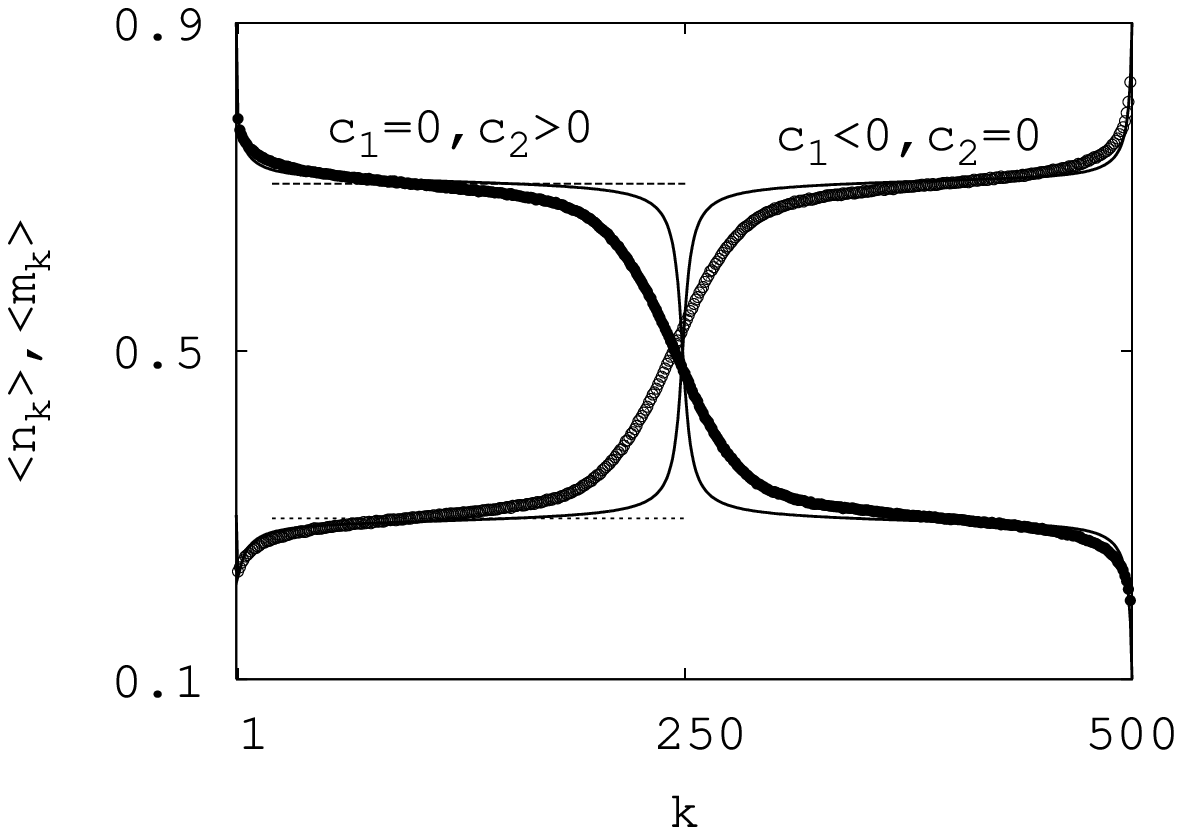}}
\qquad
\subfigure[\label{Fig_UprofileQm5}] {\includegraphics[width=0.45\textwidth]{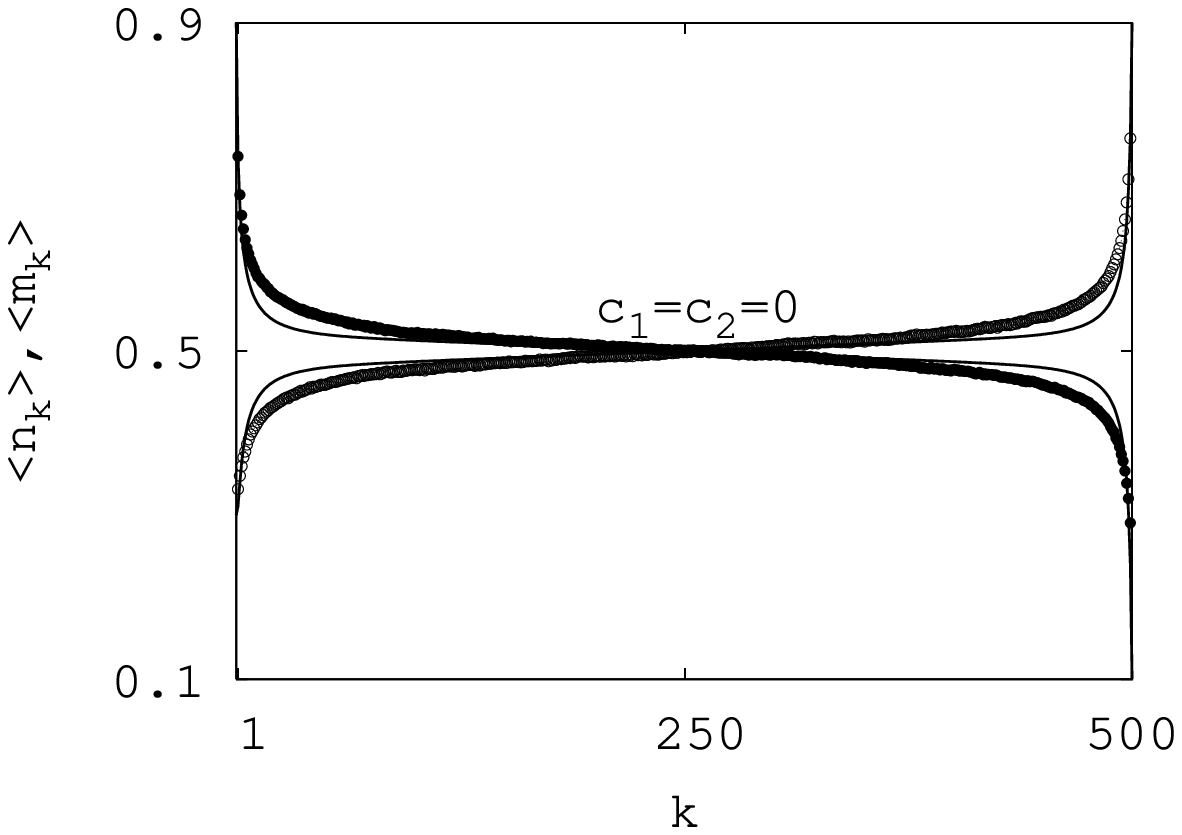}}
\end{center}
\caption{Average density profiles in phases controlled by umbilic point in a
bidirectional model Fig.\ref{Fig_BidirectionalModel}, for $Q=-0.9 < Q_{crit}$(
\textbf{Panel (a)}) and $Q=-0.5 > Q_{crit}$( \textbf{Panel (b)}). Symbols are
data points obtained from Monte Carlo simulations while lines are given by
numerical solution of hydrodynamic equations (\ref{MF}). Values of the
characteristic velocities are indicated on top of density profiles.
Parameters: $N=500$, $(u_{L},v_{L})=(0.9,0.3)$, $(u_{R},v_{R})=(0.1,0.9)$.
Note that bulk profiles are left-right symmetric while the boundary rates are
not. Broken lines on Panel (a) indicate theoretically predicted bulk values,
the U-shock amplitude being equal to $2 \Delta$ where $\Delta$ is given by
(\ref{Delta}). Both umbilic point-controlled phases are stable, if
$(u_{L},v_{L})\in G_{--}$ and $(u_{R},v_{R})\in G_{++}$, see also Tables
\ref{TableUpathPh2},\ref{TableNoUpathPh2}}%
\label{Fig_Uprofile}%
\end{figure}

An isolated umbilic point in our system (for $Q<-3/4$) was related to an
existence of a large-scale excitation, called U-shock: it is microscopically
sharp like a shock but according to usual stability criteria it should be
unstable \cite{UshockPRE2012}. The U-shock interface, see
Fig.\ref{Fig_Uprofile}(a), is connecting two rarefaction waves
\cite{StationaryRarefaction}. If the umbilic point is not isolated (for
$Q\geq-3/4 $ ), the U-shock reduces to a bulk homogeneous state with
$c_{1}=c_{2}=0$, meaning that it has densities $u^{\ast}=v^{\ast}=1/2$,
matching the umbilic point. We call this state an umbilic state. Both U-shock
and umbilic state profiles have a property of being left-right symmetric,
while the boundary conditions, generically, are not. In the next section we
describe the domain on the phase diagram, occupied by umbilic point- related
phases, and the respective phase transitions.

\section{Boundary-driven phase transitions controlled by umbilic point}

\label{sec::Boundary-driven}

The splitting of the physical region, done in the previous section, allows to
study a phase diagram of an open system, and transitions between those, in a
systematic manner. A transition from one phase to another in an open system
with fixed bulk rates happens by a gradual adiabatic change of the boundary
rates, which amounts to a respective adiabatic change of boundary densities
$u_{L},v_{L},u_{R},v_{R}$. Such a gradual change defines a path, or trajectory
in the $4$-dimensional space $\Gamma(s)\equiv$ $\{u_{L}(s),v_{L}%
(s),u_{R}(s),v_{R}(s)\}$, parametrized by a running variable $s$. Choosing the
average particle densities $u,v$ in the steady state as an order parameter, we
study a correspondence $\Gamma(s)\rightarrow(u(s),v(s))$, which shows singular
behaviour at the critical points $s_{crit}$ along a path. Across a critical
point a transition between two neighbouring phases takes place, which, see
\cite{PopkovHierarchy11} is always governed by a large scale excitation-
either a shock or a rarefaction wave, depending on how boundary changes are
performed. To specify, denote the number of positive characteristic velocities
in the left (right) boundary reservoir as $\#_{L}$ ($\#_{R}$), which can then
take integer values $\#_{R},\#_{L}=0,1,2$. A transition from one phase to
another is governed by a rarefaction wave if $\#_{L}(s)\leq\#_{R}(s)$ for all
$s$. A straightforward example of a path $\Gamma(s)$ satisfying the
requirement $\#_{L}(s)\leq\#_{R}(s)$, is a Ph2 path described by steps
I,II,III below, see also first two columns of Table \ref{TableUpathPh2}. In a
strictly hyperbolic system a mapping $\Gamma(s)\rightarrow(u(s),v(s))$ along a
Ph2 path is continuous \cite{PopkovHierarchy11}.

An isolated umbilic point, present in our model for large interlane
interactions $Q<-3/4$, stabilizes a sharp interface connecting two rarefaction
waves, a so-called a U-shock \cite{UshockPRE2012}. Below we demonstrate that
the U-shock makes possible a discontinuous change $\Gamma(s)\rightarrow
(u(s),v(s))$ along a Ph2 path in a system with an isolated umbilic point.

In order to clarify the influence of a U-shock on the phase diagram we
consider a Ph2 path \cite{Ph2path} from a steady state $G_{++}$ to a steady
state $G_{--}$, in presence and in absence of an isolated umbilic point. Let
us denote by $0\leq s\leq1$ a variable parametrizing the adiabatic Ph2 path,
and by $u_{L}(s),v_{L}(s),u_{R}(s),u_{R}(s)$ the respective densities of
boundary reservoirs. As in \ \cite{PopkovHierarchy11}, we shall vary the
boundary densities along the Ph2 path in the following way:

I. Initial point $s=0$ and final point $s=1$ corresponds to fully-matching
left and right boundary reservoirs. $u_{L}(0)=u_{R}(0)=u^{initial}%
,v_{L}(0)=v_{R}(0)=v^{initial}$, where $c_{1}(u^{initial},v^{initial}%
)>0,c_{2}(u^{initial},v^{initial})>0$. Analogously for $u_{L}(1)=u_{R}%
(1)=u^{final},v_{L}(0)=v_{R}(0)=v^{final}$, where $c_{1}(u^{final}%
,v^{final})<0,c_{2}(u^{final},v^{final})<0$.

II. For $0\leq s\leq\frac{1}{2}$, left boundary densities $u_{L}(s),v_{L}(s)$
are changing smoothly from $u^{initial},v^{initial}$ to $u^{final},v^{final}$
at $s=1/2$. The right boundary densities remain the same, $u_{R}%
(s\leq1/2)=u^{initial},v_{R}(s\leq1/2)=v^{initial}$

III. For $\frac{1}{2}\leq s\leq1$, the right boundary densities $u_{R}%
(s),v_{R}(s)$ are changing smoothly from $u^{initial},v^{initial}$ at
$s=\frac{1}{2}$to $u^{final},v^{final}$ at $s=1$. The left boundary densities
remain the same, $u_{R}(s\geq1/2)=u^{final},v_{R}(s\geq1/2)=v^{final}$.

As argued in \cite{PopkovHierarchy11}, all along such a path the steady state
is controlled by rarefaction waves. In absence of an umbilic point,
rarefaction waves do not have any discontinuities \cite{Lax2006},\cite{Lax73},
and therefore the stationary densities along the path $u(s),v(s)$ are expected
to change continuously with $s$.

In the following we demonstrate that an isolated umbilic point provokes two
discontinuous jumps of $u(s),v(s)$, and locate the critical point. The jumps
are due to a motion of the U-shock between the boundaries which \ become
biased at the transition point.

\begin{figure}[ptbh]
\begin{center}
\subfigure[\label{Fig_Qm09::a}] {\includegraphics[width=0.8\textwidth]{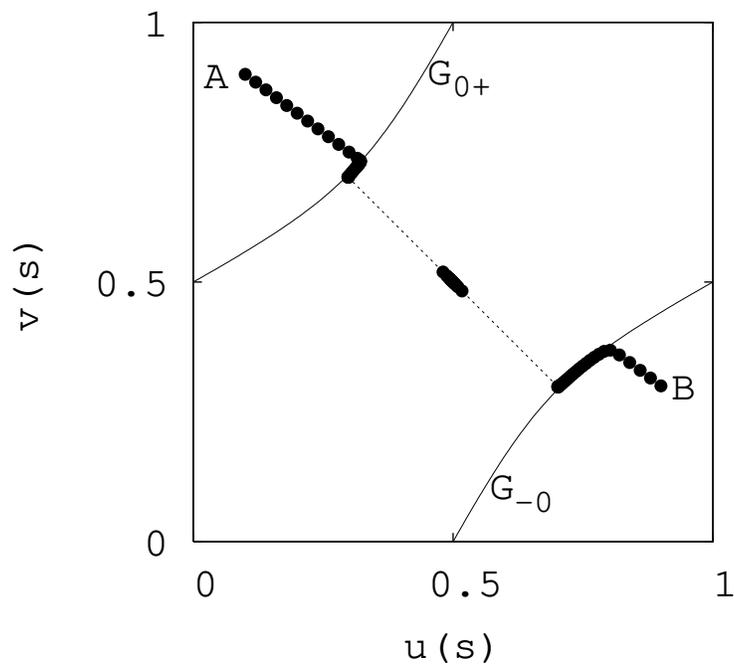}}
\qquad
\subfigure[\label{Fig_Qm09::b}]{\includegraphics[width=0.8\textwidth]{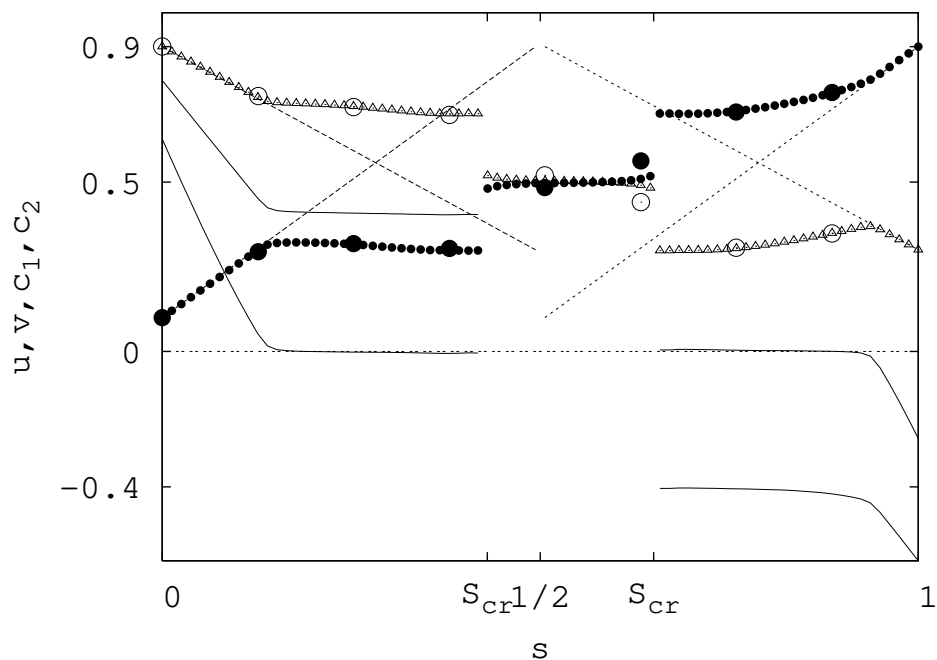}}
\end{center}
\caption{ \textbf{Panel (a)}: Location of stationary bulk densities
$u(s),v(s)$ along the Ph2 path on a $u-v$ plane, in case of isolated umbilic
point. Points A and B mark initial and final point of the path, respectively.
Curves marked $G_{0+}$ and $G_{-0}$ denote location of points with
$c_{1}(u,v)=0$ and $c_{2}(u,v)=0$, respectively. Broken line mark a
discontinuous transition to the U-shock phase. Spreading of the points in the
middle is due to finite size-effects. \textbf{Panel (b)}: Stationary bulk
densities $u(s),v(s)$ of the right movers (small circles) and of the left
movers (triangles) along the Ph2 path. Broken lines denote the dependence of
$(u_{L}(s),v_{L}(s))$ for $s<1/2$ and $(u_{R}(s),v_{R}(s))$ for $s>1/2$. Thin
lines denote $c_{k}(u,v)$. Critical points $s_{cr}$ mark discontinuous
transitions to the U-shock phase. Most data are obtained by numerical
integration of the meanfield equations. Large circles on Panel(b) mark steady
state densities obtained from Monte-Carlo simulations. Parameters: $Q=-0.9$,
$N=200$.}%
\label{Fig_Qm09}%
\end{figure}

\begin{figure}[ptbh]
\begin{center}
\subfigure[\label{Fig_Qm05::a}] {\includegraphics[width=0.8\textwidth]{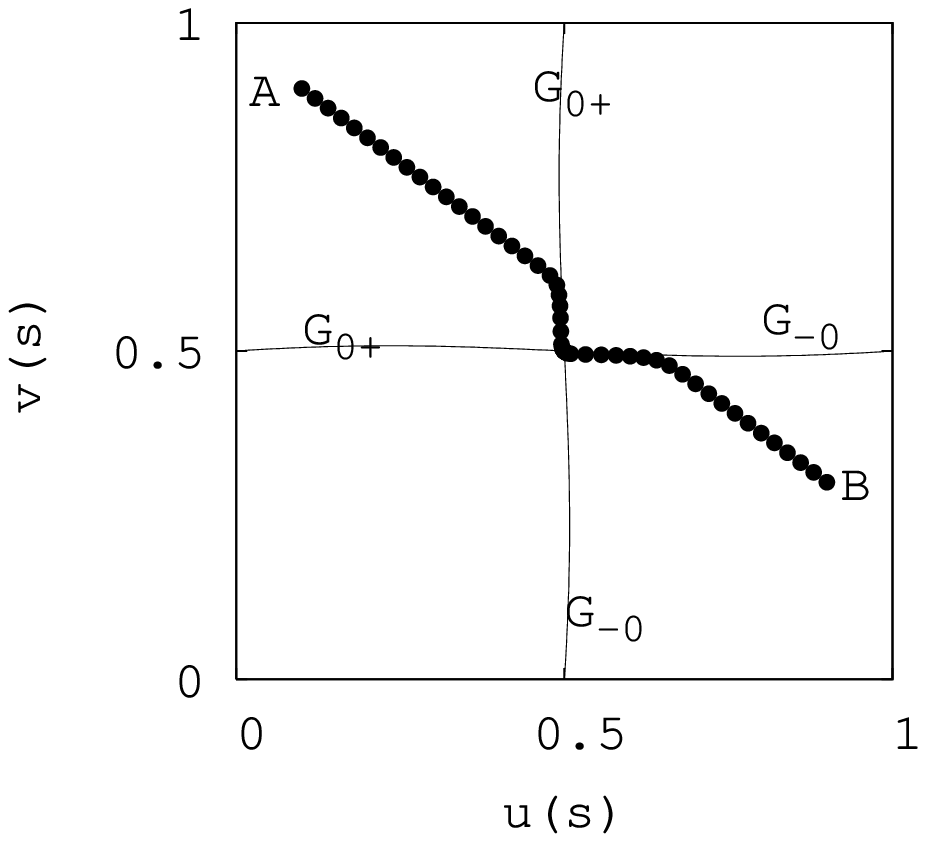}}
\qquad
\subfigure[\label{Fig_Qm05::b}]{\includegraphics[width=0.8\textwidth]{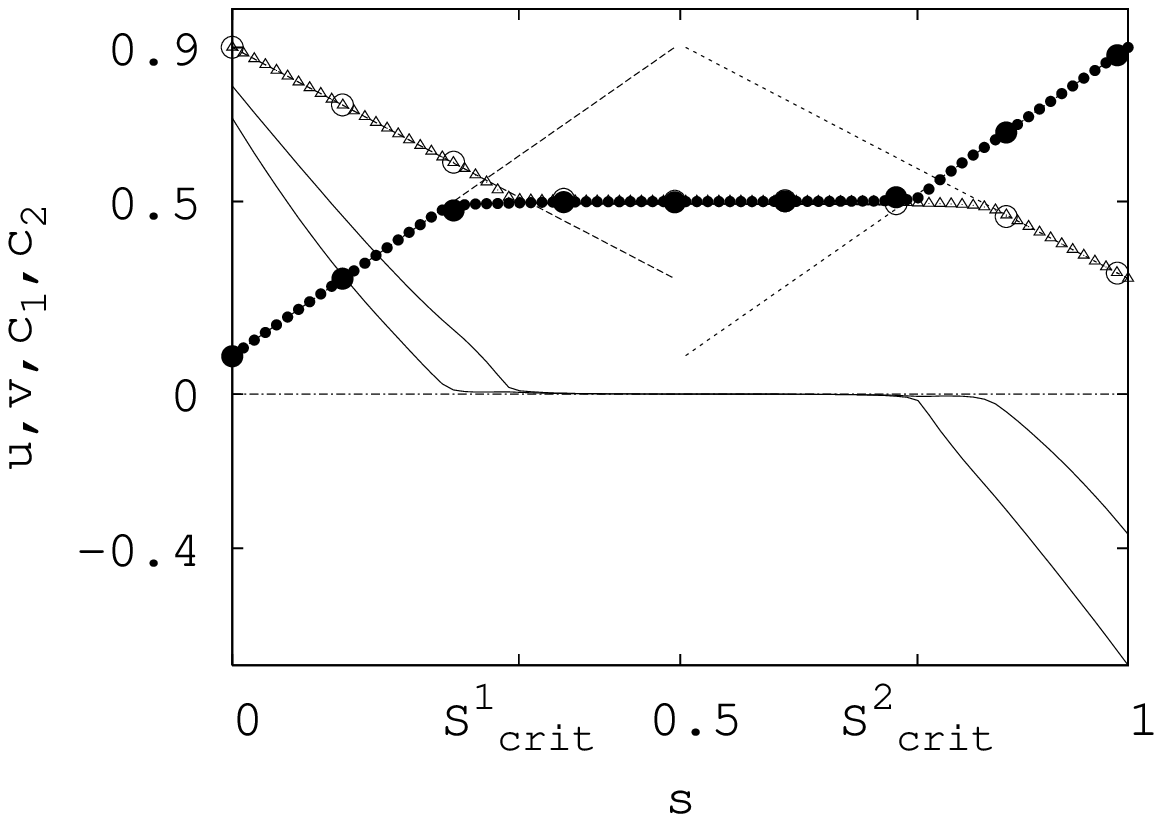}}
\end{center}
\caption{ \textbf{Panel (a)}: Location of stationary bulk densities
$u(s),v(s)$ along the Ph2 path on a $u-v$ plane, for $Q=-0.5$. Points A and B
mark initial and final point of the path, respectively. Curves marked $G_{0+}
$ and $G_{-0}$ denote location of points with $c_{1}(u,v)=0$ and
$c_{2}(u,v)=0$, respectively. \textbf{Panel (b)}: Stationary bulk densities
$u(s),v(s)$ of the right movers (circles) and of the left movers (triangles)
along the Ph2 path. Broken lines denote the dependence of $(u_{L}%
(s),v_{L}(s))$ for $s<1/2$ and $(u_{R}(s),v_{R}(s))$ for $s>1/2$. Thin lines
denote $c_{k}(u,v)$. Critical points $s_{crit}^{1}$ and $s_{crit}^{2}$ mark
transitions to the umbilic phase. Results are obtained by numerical
integration of the meanfield equations. Large circles on Panel(b) mark steady
state densities obtained from Monte-Carlo simulations. Parameters: $Q=-0.5$,
$N=200$.}%
\label{Fig_Qm05}%
\end{figure}

We choose the initial and the final state to be $(u^{initial},v^{initial}%
)=(0.1,0.9)$ and $(u^{final},v^{final})=(0.9,0.3)$, and change the boundary
densities for intermediate $s$ by linear interpolation, i.e. $u_{L}%
(s)=u^{initial}+2s(u^{final}-u^{initial}),$ $v_{L}(s)=v^{initial}%
+2s(v^{final}-v^{initial})$ for $0\leq s\leq1/2$, and similarly for $s\geq
1/2$. After that we perform an adiabatic Ph2 path for a system with an
isolated umbilic point $Q=-0.9<Q_{crit}$, and non-isolated umbilic point
$Q=-0.5<Q_{crit}$. For both values of $Q$, $(u^{initial},v^{initial})\in
G_{++}$ and $(u^{final},v^{final})\in G_{--}$. We present the results in Figs.
\ref{Fig_Qm09},\ref{Fig_Qm05}.

\begin{figure}[ptb]
\centerline{\scalebox{0.6}{\includegraphics[width=0.8\textwidth]{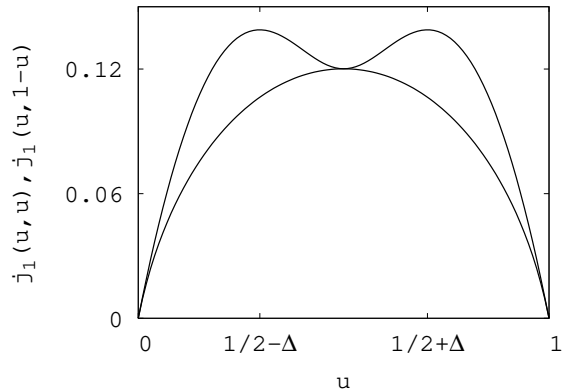}}
}\caption{Cuts of the surface $j_{1}(u,v)$ along the lines $u=v$ (the convex
curve) and $u=1-v$ (the curve with double maximum), for $Q<Q_{crit}$.
Parameters: $Q=-0.9$. Points $u=1/2\pm\Delta$ correspond to current maxima,
$\Delta$ is given by (\ref{Delta}). Above the critical point $Q>Q_{crit}$ both
cuts are convex. }%
\label{FigCuts}%
\end{figure}

\textbf{Case 1. Isolated umbilic point.} An isolated umbilic point appears
above the critical interaction amplitude, $Q<-\frac{3}{4}$. In this case, Fig.
\ref{Fig_Qm09}(a) , we see a discontinuous change in average stationary
densities $u(s)$ and $v(s)$ at two points $s_{crit}^{1},s_{crit}^{2}.$ As
already stressed, this discontinuous phase transition is impossible in a
strictly hyperbolic system. A closer examination reveals the mechanism of the
new transition. Firstly, we notice that the steady state of a system is a
U-shock if
\begin{equation}
(u_{L},v_{L})\in G_{++}\text{ \ and }(u_{R},v_{R})\in G_{--},
\label{UshockDomain}%
\end{equation}
which corresponds to a segment $s\in\lbrack s_{crit}^{1},s_{crit}^{2}]$ of the
adiabatic Ph2 trajectory. The U-shock is an interface between the two
rarefaction waves. The average bulk densities to the left and to the right
from the interface depend only on $Q$ and are given by $u,v=\frac{1}{2}%
\pm\Delta$, where
\begin{equation}
\Delta=\frac{1}{4}\sqrt{3Q^{-1}+4}\text{.} \label{Delta}%
\end{equation}
The above values of $u,v$ correspond to maxima of the current density relation
$j(u,v,Q)$, which develops a saddle point for $Q<-3/4$, see Fig.\ref{FigCuts}.
Steady-state currents of both species have the same amplitude, $j_{1}%
^{U}=-j_{2}^{U}=1/(8\left\vert Q\right\vert )$.

As long as the conditions (\ref{UshockDomain}) are satisfied, the bulk
densities, and consequently the steady currents, do not depend on boundary
densities. The boundary densities affect only the respective boundary layers,
interpolating between the bulk and the boundary, on a microscopic scale. The
interface position fluctuates with time around the center of the lattice.

Now consider a left vicinity of a critical point $s=s_{crit}^{1}-\varepsilon$,
where $0<\varepsilon\ll1$. Once $\varepsilon>0$, the U-shock is biased to the
left, and gets pinned to the left boundary. The resulting steady state is
homogeneous with the densities $u=\frac{1}{2}-\Delta,v=\frac{1}{2}+\Delta$.
With $\varepsilon$ increasing, $(u(s),v(s))$ follows the curve $G_{+0}$.

Analogously, at the other side of the U-shock phase, in the right vicinity of
another critical point $s=s_{crit}^{2}+\varepsilon$, the U-shock is biased to
the right. This results in a homogeneous steady state with the densities
$u=\frac{1}{2}+\Delta,v=\frac{1}{2}-\Delta$, where $\Delta$ is given by
(\ref{Delta}). As $s$ increases, the point $(u(s),v(s))$ follows the curve
$G_{0-}$. Depinning from the curve $G_{0-}$ occurs when both $(u_{L},v_{L})\in
G_{++}$ \ and $(u_{R},v_{R})\in G_{++}$.

Steady states and corresponding reservoir densities are listed in order of
their temporal appearance along the adiabatic trajectory on Table
\ref{TableUpathPh2}. As one can see, steady states are straightforwardly
connected to the phase space splitting in subregions $G_{\alpha\beta}$ with
different signs of characteristic velocities.

\bigskip

\begin{table}[ptb]
\vspace{0.5cm}
\begin{tabular}
[c]{c|c|c|c}\hline
$(u_{L},v_{L}) $ & $(u_{R},v_{R})$ & Steady state & Steady state\\
belong to & belong to & belongs to & densities\\\hline
$G_{++}$ & $G_{++}$ & $G_{++}$ & $u=u_{L}$, $v=v_{L}$\\
$G_{-+}$ & $G_{++}$ & $G_{0+}$ & pinned to $G_{0+}$\\
$G_{--}$ & $G_{++}$ & $G_{0+}$/$G_{-0}$ & U-shock $u=\frac{1}{2} \pm\Delta$,
$v=\frac{1}{2} \mp\Delta$\\
$G_{--}$ & $G_{-+}$ & $G_{-0}$ & pinned to $G_{-0}$\\
$G_{--}$ & $G_{--}$ & $G_{--}$ & $u=u_{R}$, $v=v_{R}$\\\hline
\end{tabular}
\par
\caption{Steady states densities $u(s),v(s)$ and corresponding particle
densities in boundary reservoirs along a Ph2 path. $\Delta$ is given by
(\ref{Delta}). }%
\label{TableUpathPh2}%
\end{table}

\begin{table}[ptb]
\vspace{0.5cm}
\begin{tabular}
[c]{c|c|c|c}\hline
$(u_{L},v_{L}) $ & $(u_{R},v_{R})$ & Steady state & Steady state\\
belong to & belong to & belongs to & densities\\\hline
$G_{--}$ & $G_{++}$ & $G_{00}$ & umbilic point rarefaction $u=v=\frac{1}{2}%
$\\\hline
\end{tabular}
\par
\caption{Open bidirectional system with a non-isolated umbilic point. Steady
states and corresponding boundary reservoir densities along the middle part of
a Ph2 path. For remaining path segments, see Table \ref{TableUpathPh2}}%
\label{TableNoUpathPh2}%
\end{table}It is instructive to compare the discontinuous U-shock-governed
phase transition described above to a usual discontinuous phase transition in
driven systems governed by a standard Lax shock \cite{Lax73}. In the latter,
the shock changes the sign of a bias at the critical point. Exactly at the
critical point, the shock is unbiased, and, due to fluctuations, performs a
random walk between the boundaries. On the contrary, in a U-shock --controlled
phase transition, the U-shock stays unbiased in the whole segment
$[s_{crit}^{1},s_{crit}^{2}]$ , defined by (\ref{UshockDomain}). For $s$
within this segment, the U-shock position is not performing a random walk
between the boundaries, but fluctuates around the middle point as if it was in
a potential well.

\textbf{Case 2. Non-isolated umbilic point. }$\ $\ Now, let us push the
interaction $Q$ towards the critical point $Q_{crit}=-3/4$. The U-shock
amplitude $2\Delta$ becomes smaller and smaller until it disappears at the
critical point $\Delta(Q_{crit})=0$. The whole U-shock--governed phase above
the critical point $Q\geq Q_{crit}$ reduces to a homogeneous phase with
densities matching the umbilic point $c_{1}(u,v)=c_{2}(u,v)=0$, see Table
\ref{TableNoUpathPh2}. In this way, the umbilic point defines a robust phase
on the phase diagram, which we shall call an umbilic phase or U-phase. The
U-phase with $c_{1}=c_{2}=0$ appears to be stable whenever $(u_{L},v_{L})\in
G_{--}$ and $(u_{R},v_{R})\in G_{++}$, see Table \ref{TableNoUpathPh2}, in the
very same domain where a U-shock is stable (\ref{UshockDomain}). A
pinning-depinning transition from the U-phase is continuous, in contrary to a
transition from/to U-shock --governed phase for $Q<Q_{crit}$ discussed
earlier, see Fig.\ref{Fig_Qm05}.

Several comments are in order at this point. Firstly, note that the steady
state particle currents and bulk density profiles for an umbilic phase and for
a U-shock phase are left-right symmetric (so as the bulk hopping rates), in
spite of boundary conditions being explicitly not left-right symmetric. Away
from the umbilic (or U-shock) phase, a steady state is sensitive to the
boundaries and in general is not left-right symmetric. In this way, when
entering the respective U-phases, the steady state becomes insensitive to
boundaries and regains its bulk symmetry. In absence of umbilic points (i.e.
in strictly-hyperbolic systems), steady states are generically boundary-rates dependent.

Secondly, we observe that the total current of both species $j_{1}+\left\vert
j_{2}\right\vert $ along the Ph2 path attains its maximal value just for an
umbilic phase and of a U-shock phase. This fact exemplifies a validity of the
maximal current principle for bidirectional systems. Indeed, in systems with
one driven particle species and open boundaries the maximal current principle
asserts \ in particular that for maximal feeding regime (when both entrance
and exit rates are maximal), the stationary current is maximized with respect
to an average particle density $\rho$, $j_{steady}=\max_{\rho\in\lbrack
0,\rho_{\max}]}j(\rho)$. In our model with two species, such a maximal feeding
regime is realized when $(u_{L},v_{L})=(1,0)$ and $(u_{R},v_{R})=(1,0)$,
which, consulting the Fig.\ref{FigSplitting} and (\ref{UshockDomain}),
corresponds to the U-shock or Umbilic phase domain, depending on the value of
$Q$. The respective maximal value of the total current is $j_{1}+\left\vert
j_{2}\right\vert =2j_{1}=(4\left\vert Q\right\vert )^{-1}$ for U-shock
$Q<Q_{crit}$ and $j_{1}+\left\vert j_{2}\right\vert =2j_{1}(u=\frac{1}%
{2},v=\frac{1}{2})=\sqrt{Q+1}/(\sqrt{Q+1}+1)$ for the umbilic phase $Q\geq
Q_{crit}$, as can be straighforwardly derived from analytic expressions for
the currents.

\section{Conclusions}


We conclude that a presence of an umbilic point in the current-density
function, isolated or not, gives rise to new types of boundary driven phase
transitions. If the umbilic point is an isolated one, one has a discontinuous
transition from $G_{0+}$ steady state to the U-shock state $G_{0+}/G_{-0}$ and
another discontinuous transition from the U-shock state to $G_{-0}$ state,
both transitions governed by a biased motion of a U-shock. In case of
non-isolated umbilic point, two continuous (pinning-depinning) transitions
$G_{0+}\rightarrow G_{00}$ and $G_{00}\rightarrow G_{-0}$ take place. All
these transitions are observable along any Ph2 path, leading from a state with
positive characteristics to a steady state with negative characteristics via
adiabatically changing boundary conditions. There is no hysteresis of any
kind, so by inversion of the path the sequence of transitions is inverted. The
U-shock phase and umbilic phases are robust, and exemplify the validity of a
maximal current principle for bidirectional particle models. We identified the
domain on the phase diagram occupied by umbilic point related phases
(\ref{UshockDomain}). Within this domain the system regains its bulk
left-right symmetry in spite of the boundary conditions being explicitly not
left-right symmetric.

How robust are our results? First of all, the umbilic point with $c_{1}%
=c_{2}=0$ is a general feature of a models with left-right symmetry of hopping
rates \cite{UshockPRE2012}, of which we considered a special example with an
exactly known steady state on a ring. We expect qualitatively similar results
for general particle systems whether "solvable" or not. In an open system it
is only necessary to maintain a stationary maximal flow regime to find out
whether the umbilic point is isolated or not. Appearance of a discontinuity in
a bulk density profile in the maximal flow regime would indicate that an
umbilic point has become isolated. In such a way from a simple single
macroscopic observation of a system one can judge on intrinsic differential
properties of its current-density \ function.

Bidirectional models are being widely studied in the literature, in
particular, in connection with the intriguing phenomenon of spontaneous
symmetry breaking (SSB) \cite{bridge}-\cite{bridgeJiang}. In most studies
however the current-density function is a convex surface. It would be
interesting to study SSB in presence of an isolated umbilic point. It is
interesting to note that bidirectional models have also an integrability
aspect \cite{PopkovSchutzEbrahim02}, however up to now no example of an
integrable system with open nontrivial boundary conditions have been
presented.

\section*{Acknowledgements}

V.P. thanks the IZKS and the University of Bonn for hospitality and
acknowledges a partial support by the Alexander von Humboldt foundation, and
by the italian MIUR through PRIN 20083C8XFZ initiative.

\appendix

\section{Boundary rates}

\label{Appendix::Boundary rates}

Boundary rates for injection and extraction of the particles are chosen so
that constant densities of particles are kept on the left and on the right
boundary. Steady state of our model with the rates (\ref{rates}) for a system
on a ring has a remarkable property: for a configuration $C$ the stationary
probability is given by a product measure
\begin{equation}
P_{C}=Z^{-1}%
{\displaystyle\prod\limits_{k}}
e^{-\nu n_{k}m_{k}}, \label{ProductMeasure}%
\end{equation}
where $n_{k},m_{k}=0,1$ are particle occupation number on site $k$ on chains
$1$ and $2$, and $Z$ is a normalization. We see that neighbouring pairs of
adjacent sites are uncorrelated. This fact allows to express all steady state
equal-time expectations in terms of probabilities of a single pair of adjacent
sites $\Omega_{n_{k}m_{k}}(u,v)$ where $u,v$ is an average density of right
and left movers. We have $\Omega_{00}+\Omega_{11}+\Omega_{01}+\Omega_{10}=1$,
$\Omega_{10}+\Omega_{11}=u$, $\ \Omega_{01}+\Omega_{11}=v$, and $\Omega
_{11}(u,v)$ is given by (\ref{Omega11}). Procedure, completely analogous to
that \ in \cite{PopkovPeschel} results in the following definition of the
boundary rates: a right-moving particle is injected to a site $1$ with rate
$u_{L}$ (with rate $u_{L}+Q\Omega_{11}(u_{L},v_{L})$ ) if an adjacent site is
empty (filled). A right-moving particle is extracted from site $N$ with rate
$1-u_{R}$ (with rate $1-u_{R}+Q\Omega_{00}(u_{R},v_{R})$ ) if an adjacent site
is filled (empty). Analogously, left moving particles are injected at the
right boundary with rate $v_{R}$ (with rate $v_{R}+Q\Omega_{11}(v_{R},u_{R})$
) if an adjacent site is empty (filled), and extracted at the left boundary
with rate $1-v_{L} $ (with rate $1-v_{L}+Q\Omega_{00}(v_{L},v_{L})$ ) if an
adjacent site is filled (empty). In case of matching left and right
boundaries, $u_{L}=u_{R}=u,v_{L}=v_{R}=v$, the exact steady state of the
system is (\ref{ProductMeasure}), independently of system size $N$.

\section{Meanfield equations}

\label{Appendix::Meanfield equations}

In our meanfield approximation, we neglect correlations between the adjacent
pairs of sites, which are also absent in the steady state
(\ref{ProductMeasure}) but not between the adjacent sites. The equations are
obtained by averaging the exact operator equations of motion for occupation
number operators $\hat{n}_{k},\hat{m}_{k}$, which for right-moving particle at
site $k$ read
\[
\frac{\partial\langle n_{k}\rangle}{\partial t}=\langle\hat{\jmath}%
_{k-1,k}\rangle-\langle\hat{\jmath}_{k,k+1}\rangle
\]
where $\hat{\jmath}_{k,k+1}=\langle\hat{n}_{k}(1-\hat{n}_{k+1})\rangle
+Q\langle\hat{n}_{k}\hat{m}_{k}(1-\hat{n}_{k+1})(1-\hat{m}_{k+1})\rangle$.
Denoting $\langle\hat{n}_{k}(t)\rangle=s_{k}(t)$, $\langle\hat{m}%
_{k}(t)\rangle=q_{k}(t)$, we simplify parts of above expression as
$\langle\hat{n}_{k}(1-\hat{n}_{k+1})\rangle\approx\langle\hat{n}_{k}%
\rangle\langle1-\hat{n}_{k+1}\rangle=s_{k}(1-q_{k+1})$ and $\langle\hat{n}%
_{k}\hat{m}_{k}(1-\hat{n}_{k+1})(1-\hat{m}_{k+1})\rangle\approx\langle\hat
{n}_{k}\hat{m}_{k}\rangle\langle(1-\hat{n}_{k+1})(1-\hat{m}_{k+1}%
)\rangle=\Omega_{11}(s_{k},q_{k})\Omega_{00}(s_{k+1},q_{k+1})$, where
$\Omega_{11}(u,v)$ is given by exact microscopic expression (\ref{Omega11}).
Thus, for a homogeneous state $\langle\hat{n}_{k}\rangle=u$, $\langle\hat
{m}_{k}\rangle=v$ the meanfield expression gives the exact microscopic
stationary current (\ref{Ju}). The semiclassical equation of motion becomes%

\begin{align}
\frac{\partial s_{k}}{\partial t}  &  =+s_{k-1}(1-q_{k})-Q\Omega_{11}%
(s_{k-1},q_{k-1})\Omega_{00}(s_{k},q_{k})\nonumber\\
&  -s_{k}(1-q_{k+1})-Q\Omega_{11}(s_{k},q_{k})\Omega_{00}(s_{k+1},q_{k+1}),
\label{MF}%
\end{align}
for $k=2,3,...N-1$, complemented with boundary conditions for the boundary
sites $k=1$ and $k=N$%
\begin{equation}
\text{\ }s_{1}(t)=u_{L};\text{ \ }s_{N}(t)=u_{R};\text{ } \label{MF_BC}%
\end{equation}
For left-movers the equations are derived analogously. The complete set of
equations of motion can be viewed as a natural discretization scheme with
which we integrate numerically the hydrodynamic equations (\ref{PDE}). Indeed,
by Taylor expansion of (\ref{MF}), and Euler rescaling of space and time we
obtain (\ref{PDE}). Comparison with the stochastic evolution shows that both
steady state profiles and temporal evolution is described correctly.


\begin{thebibliography}{99}
\bibitem {DerivationHydro}The hydrodynamic limit equations can be obtained by
averaging operator equations of motion, as in Appendix
\ref{Appendix::Meanfield equations}, consequent Taylor expansion $\langle
n_{k+1}(t)\rangle\rightarrow u(x+\frac{a}{N},t)\approx u(x,t)+\frac{a}%
{N}\partial u/\partial x$, etc., and Euler rescaling of space and time.
\bibitem {StationaryRarefaction}A stationary rarefaction wave is characterized
by the fact it has one zero characteristic velocity ($c_{i}=0$), as opposed to
a stationary shock wave where all characteristic velocities are nonzero
\cite{Lax73}.
\bibitem {Ph2path}We borrowed the notation from \cite{PopkovHierarchy11}.
\end{thebibliography}

\end{document}